\documentclass[aps,prc,preprint,groupedaddress,preprintnumbers,12pt]{revtex4-1}
\usepackage{natbib}
\usepackage{amsmath}
\usepackage{graphicx}
\usepackage{hyperref}
\usepackage{slashbox}
\usepackage{longtable}
\usepackage{bm}
\newcommand{\physrep}{Physics Reports}
\begin{document}

\preprint{JLAB-THY-13-1712}

\title{$^2$H$(e,e'p)$ observables using a Regge model parameterization of final state interactions}

\author{William P. Ford$^{(1)}$}
\email[]{wpford@jlab.org}
\author{Sabine Jeschonnek$^{(2)}$}
\email[]{jeschonnek.1@osu.edu}
\author{J. W. Van Orden$^{(1,3)}$}
\email[]{vanorden@jlab.org}
%\thanks{}
\affiliation{\small \sl (1) Department of Physics, Old Dominion University, Norfolk, VA
23529\\ (2) The Ohio State University, Physics
Department, Lima, OH 45804 \\ and\\ (3) Jefferson Lab\footnote{Notice: Authored by Jefferson Science Associates, LLC under U.S. DOE Contract No. DE-AC05-06OR23177.
The U.S. Government retains a non-exclusive, paid-up, irrevocable, world-wide license to publish or reproduce this manuscript for U.S. Government purposes},
12000 Jefferson Avenue, Newport News, VA 23606}
\date{\today}
\begin{abstract}
In previous papers we have presented a calculation describing electrodisintegration of the deuteron at GeV energies. The model is fully relativistic and incorporates full spin dependence of the final state interactions (FSI), which were obtained from the SAID analysis. It was, however, limited kinematically due to lack of availability of the SAID amplitudes. This work rectifies this problem by implementing a Regge model to describe the FSI. We present an outline of the model and show comparisons between the two approaches in a region of overlap. We see good agreement between the models, and note observables which can provide additional insight due to model sensitivity.
\end{abstract}
\maketitle

\section{Introduction}

In previous papers we introduced a new model of deuteron electrodisintegration for use at $Q^2>1\ {\rm GeV^2}$. The model is based on the Bethe-Salpeter equation and uses the distorted-wave impulse approximation (DWIA) with the bound state vertex function used in previous calculations of elastic electron scattering from the deuteron and a one-body current of the usual Dirac-plus-Pauli form. The inclusion of final state interactions used in this model contain all possible spin-dependence of the $np$ scattering matrix unlike previous calculations intended for use at large momentum transfers. Calculations were presented for observables for unpolarized hadrons \cite{JVO_2008_newcalc}, for deuteron polarization \cite{JVO_2009_tar_pol} and for polarization of the ejected proton \cite{JVO_2009_ejec_pol}.

The final state interactions are included in the model by means of the Fermi invariant parameterization
\begin{align}
   \hat{M} &= {{\cal F}_{S}(s,t)}1^{(1)}  1^{(2)}+ {{\cal F}_V(s,t)} \gamma^{\mu(1)} \gamma_{\mu}^{(2)}
           + {{\cal F}_T(s,t)} \sigma^{\mu \nu (1)} \sigma_{\mu \nu}^{(2)} \nonumber \\
           &- {{\cal F}_P(s,t)} (i\gamma_5)^{(1)}  (i\gamma_5)^{(2)}
           + {{\cal F}_A(s,t)} (\gamma_5\gamma^{\mu})^{(1)} (\gamma_5\gamma_{\mu})^{(2)} \label{eq:Fermi}
\end{align}
which includes all possible independent spin structures for onshell $NN$ scattering. In the previous work, the invariant functions ${\cal F}_i(s,t)$ were obtained by calculating the five independent helicity matrix elements of (\ref{eq:Fermi}), inverting these equations to obtain the ${\cal F}_i(s,t)$ in terms of the helicity matrix elements and using the $np$ helicity matrix elements available from the SAID partial wave analysis \cite{SAIDdata,SAIDpaper}. Since the SAID helicity amplitudes are only reliable up to a laboratory kinetic energy of 1.3 GeV, this placed limits on the kinematics that could be reached by the calculations.

In order to eliminate this limitation, we have recently completed a fit to available $NN$ cross sections and spin observables from $s=5.4\ {\rm GeV^2}$ to $s=4000\ {\rm GeV^2}$ using a model based on Regge theory \cite{FVO_Reggemodel}. The purpose of this paper is to compare the results for deuteron electrodisintegration observables using the final state interactions based on the SAID and Regge model versions of the final state interactions in a kinematic region where the two methods overlap.

In the next section we provide an outline of the production of the Fermi invariants from the Regge model. The following section contains the results of the calculations of the electrodisintegration observables and a discussion of these results for the two final state interaction models. The last section contains conclusions based on these calculations.

\section{The Regge model}

There is an extensive literature describing the origins and applications of Regge theory. We provide a short summary of this approach as used in \cite{FVO_Reggemodel} for those who may not be familiar with material.  A more extensive introduction can be found in \cite{martin,barone2002high,collins,perl,irving_Regge_phenom}.

\begin{figure}
    \centerline{\includegraphics[width=6in]{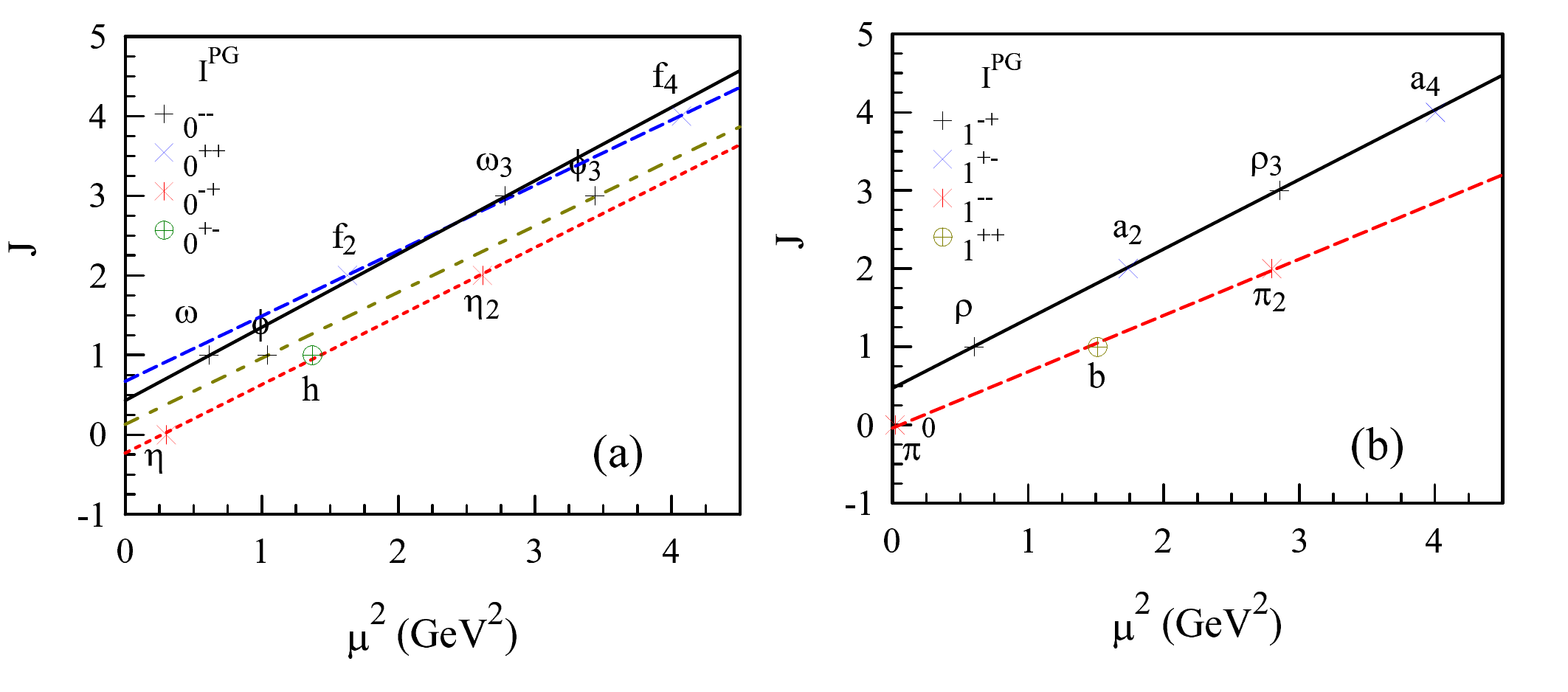}}
    \caption{Isoscalar (a) and isovector (b) mesons with spin $J$ plotted versus the square of the meson masses $\mu$. The various lines correspond to the Regge trajectories used in the fit to $NN$ scattering. An additional trajectory, the Pomeron, with $0^{++}$ is required to fit the large $s$ data.  It has an intercept of 1.08 and a slope of 0.25 GeV$^2$.}\label{fig:trajectories}
\end{figure}

The motivation for Regge theory in the case of $NN$ scattering is the observation that mesons with fixed parity, G-parity and isospin when plotted for the spin J versus the square of mass appear to lie along smooth curves as shown in Fig. \ref{fig:trajectories}. The curves can be represented as
\begin{equation}
J=\alpha(\mu^2)\,
\end{equation}
where $\alpha(\mu^2)$ is some function of the square of the meson mass $\mu^2$.  In the case of the well established mesons, the function is consistent with a straight line. The interpolating functions $\alpha_i$ describe Regge trajectories. Regge theory describes the $NN$ scattering amplitudes in terms of the exchange of Regge trajectories rather than individual mesons.

\begin{figure}
    \centerline{\includegraphics[width=3in]{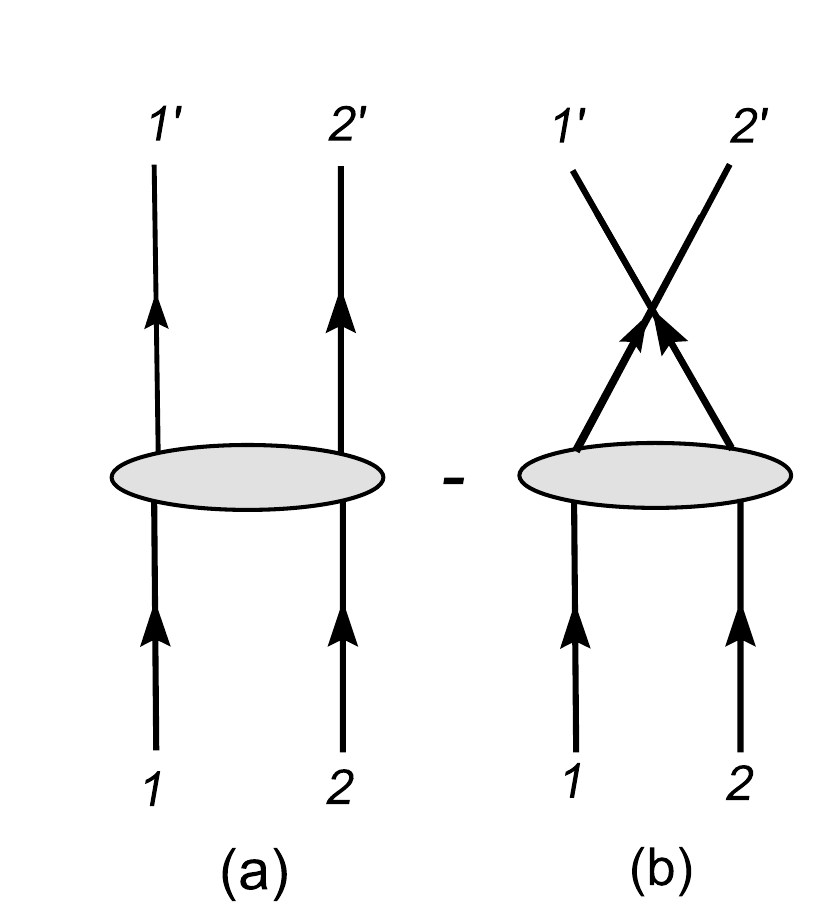}}
    \caption{(color online) Pictorial representation of the direct and exchange contributions to the $NN$ scattering amplitude.}
    \label{fig:symmetry}
\end{figure}

For $NN$ scattering the application of Regge theory can be simplified by noting that the total scattering amplitude can be described in terms of a direct contribution, Fig. \ref{fig:symmetry}(a), and an exchange contribution Fig. \ref{fig:symmetry}(b), where any set of possible diagrams represented by the ellipse is the same in both cases. This is due to the fact that any exchange of nucleons in the summation scattering diagrams can always be ``unwound'' to give contributions involving the exchange of nucleons in the final state. For this reason, we can concentrate on Fig. \ref{fig:symmetry}(a) since Fig. \ref{fig:symmetry}(b) can be obtained by the replacement of Mandelstam $t$ with $u$ and a mixing of amplitudes due to the interchange of spins in the final state.
The scattering operator associated with Fig. \ref{fig:symmetry}(a) can be written in Fermi invariant form as
\begin{align}
   \hat{M}^{(a)} &= {F_{S}^I(s,t)}1^{(1)}  1^{(2)}+ {F_V^I(s,t)} \gamma^{\mu(1)} \gamma_{\mu}^{(2)}
           + {F_T^I(s,t)} \sigma^{\mu \nu (1)} \sigma_{\mu \nu}^{(2)} \nonumber \\
           &- {F_P^I(s,t)} (i\gamma_5)^{(1)}  (i\gamma_5)^{(2)}
           + {F_A^I(s,t)} (\gamma_5\gamma^{\mu})^{(1)} (\gamma_5\gamma_{\mu})^{(2)}\,.\label{eq:Fermi_a}
\end{align}

\begin{figure}
    \centerline{\includegraphics[height=3in]{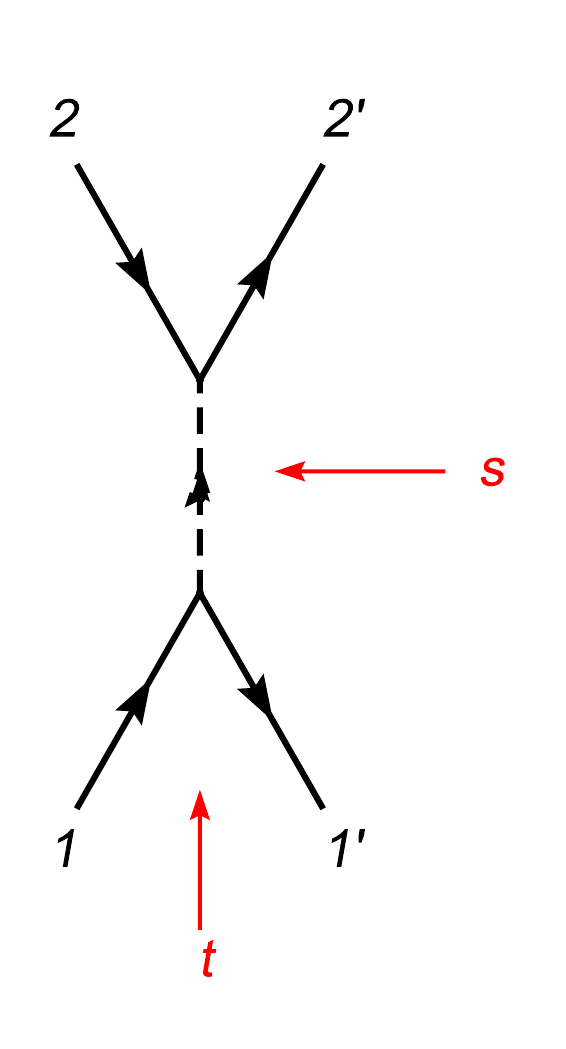}}
    \caption{(color online) Representation of the direct contribution to $N\bar{N}$ scattering through a single meson exchange in $t$-channel center of momentum frame.}
    \label{fig:tcm}
\end{figure}

Now consider a contribution to Fig. \ref{fig:symmetry}(a) from the exchange of a single meson. The meson has definite values of parity (P), G-parity (G), isospin (I) and spin (J). The propagator for the exchanged meson is a function of $t$, and since in the s-channel center of momentum (cm) frame $t<0$, the pole in the propagator can not be reached. For the purposes of Reggeization, it is useful to consider this contribution in the t-channel cm frame as represented by Fig. \ref{fig:tcm}.
This represents $N\bar{N}$ annihilating to produce a meson which then converts back to an $N\bar{N}$ state. In this frame $t>4 m^2$ and $s<0$. So for mesons of sufficiently large mass, it is possible to reach the pole in the propagator. The contribution of this diagram can be expanded in the partial wave series in the helicity basis.
\begin{equation}
M^{(a)IPG}_{\lambda_2',\lambda_2,\lambda_1',\lambda_1}=\sum_{J=0}^\infty(2J+1)\left[ f^{IGJ}_{\lambda_2',\lambda_2,\lambda_1',\lambda_1}(t)-P(-1)^{J+\lambda_1'-\lambda_1}f^{IGJ}_{-\lambda_2',-\lambda_2,\lambda_1',\lambda_1}(t)\right] d^J_{\lambda_2'-\lambda_2,\lambda_1'-\lambda_1}(\theta_t)\,,\label{eq:t_partial_wave}
\end{equation}
where the $\lambda_i$ are the helicities of the respective nucleons, $f^{IGJ}_{\lambda_2',\lambda_2,\lambda_1',\lambda_1}(s)$ is the scattering amplitude in the helicity basis for a given $I$, $P$, $G$ and $J$.

We next construct initial and final spinor states of good parity, G-parity and isospin and use these to produce matrix elements of the Fermi invariant expression (\ref{eq:Fermi_a}). Comparison of this result shows that only helicity amplitudes with $\lambda_2'=\lambda_2$ and $\lambda_1'=\lambda_1$ are required to obtain the Fermi invariants $F^{I}_i(s,t)$. So
\begin{equation}
M^{(a)IPG}_{\lambda_2,\lambda_2,\lambda_1,\lambda_1}(s,t)=\sum_{J=0}^\infty(2J+1)\left[ f^{IPGJ}_{\lambda_2,\lambda_2,\lambda_1,\lambda_1}(t)-P(-1)^{J}f^{IPGJ}_{-\lambda_2,-\lambda_2,\lambda_1,\lambda_1}(t)\right] P_J(z)\,,\label{eq:t_partial_wave_2}\,
\end{equation}
where we have used $d^J_{00}(\theta_t)=P_J(z)$ and
\begin{equation}
z=\cos\theta_t=\frac{2s}{4m^2-t}-1\,.
\end{equation}
Furthermore, the Fermi invariants can be written in terms of the particular combinations of helicity amplitudes defined as
\begin{align}
M^{(a)IPG}_\pm(s,t)&=M^{(a)IPG}_{\frac{1}{2},\frac{1}{2},\frac{1}{2},\frac{1}{2}}(s,t)\mp PM^{(a)IPG}_{-\frac{1}{2},-\frac{1}{2},\frac{1}{2},\frac{1}{2}}\nonumber\\
&=\sum_{J=0}^\infty(2J+1)\left[ f^{IPGJ}_{\frac{1}{2},\frac{1}{2},\frac{1}{2},\frac{1}{2}}(t)\mp Pf^{IPGJ}_{-\frac{1}{2},-\frac{1}{2},\frac{1}{2},\frac{1}{2}}(t)\right]\left[1\pm (-1)^J\right] P_J(z)\nonumber\\
&=\sum_{J=0}^\infty(2J+1)f^{IPGJ}_{\pm}(t)\left[1\pm (-1)^J\right] P_J(z)
\end{align}
Note that all of the dependence on $s$ is contained in the Legendre polynomial through the definition of $z$.

\begin{figure}
    \centerline{\includegraphics[height=3in]{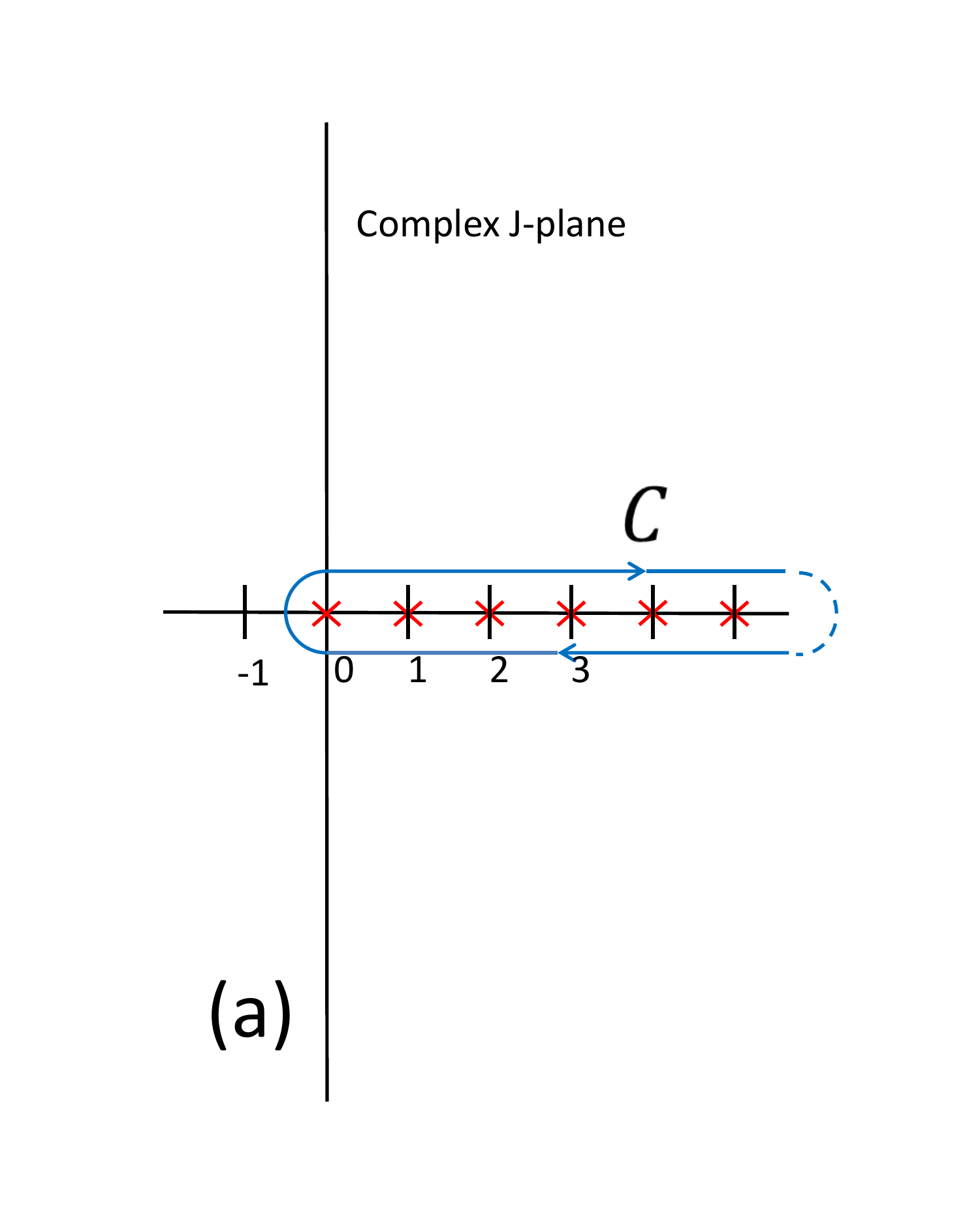}\includegraphics[height=3in]{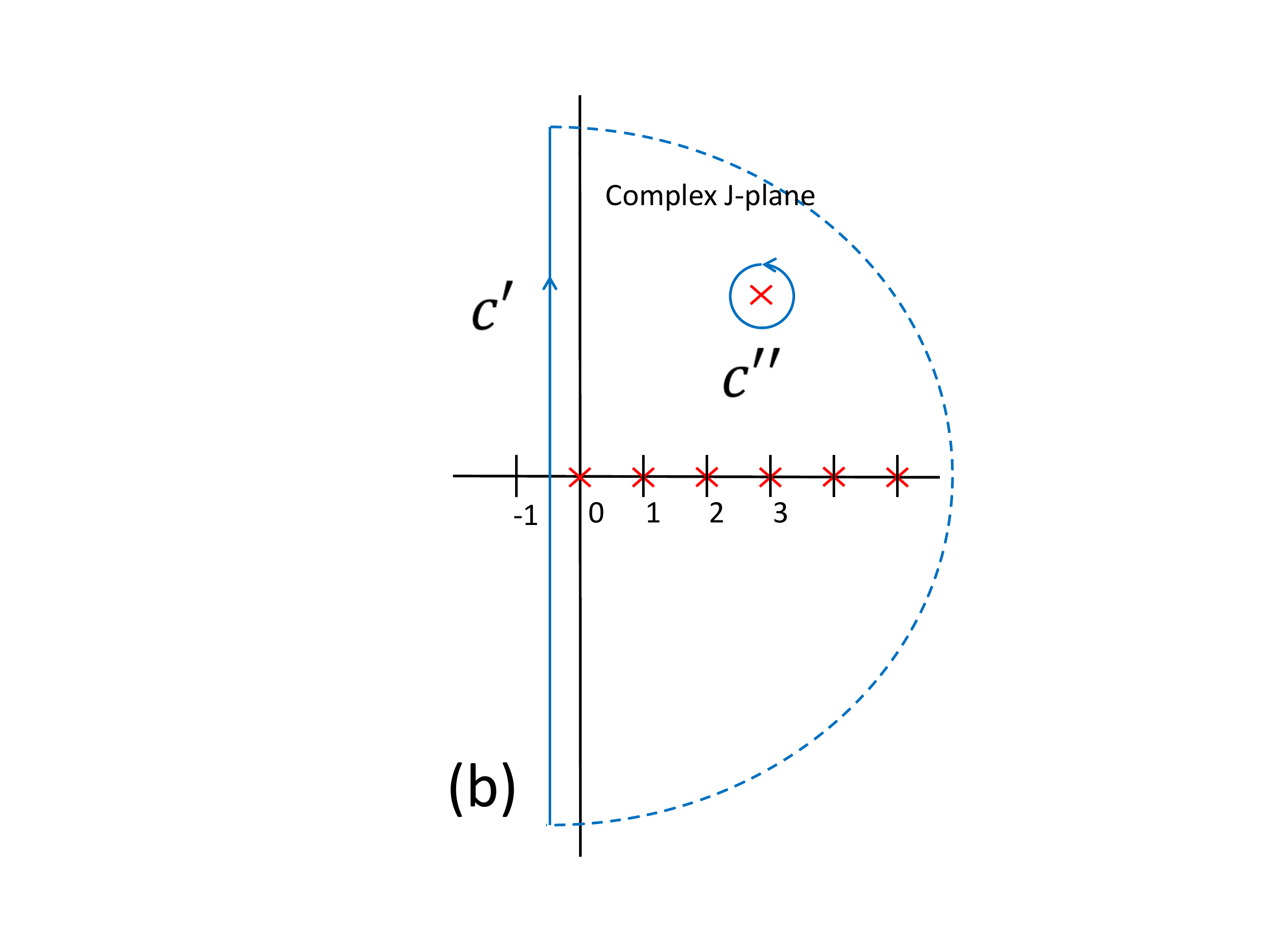}}
    \caption{(color online) Integration contours used in Reggeizing the scattering amplitudes. Diagram (a) represents the original contour used to replace the original sum over integer $J$ with an integral over complex $J$. The dashed line indicates that the contour is closed at infinity. Diagram (b) shows the distorted contour which picks a Regge pole through contour $\cal C''$. }
    \label{fig:Complex_J}
\end{figure}

The scattering matrix can now be Reggeized by replacing the sum in (\ref{eq:t_partial_wave_2}) by performing an integral over the contour $C$, as shown Fig.  \ref{fig:Complex_J}(a), to give
\begin{align}
M^{(a)IPG}_{\pm}&=-\frac{1}{2i}\oint_C dJ \frac{(-1)^J(2J+1)f^{IPG}_{\pm}(J,t)\left[1\pm (-1)^J\right] P_J(z)}{\sin\pi J}\nonumber\\
&=-\frac{1}{2i}\oint_C dJ \frac{(2J+1)f^{IPG}_{\pm}(J,t)\left( P_J(-z)\pm P_J(z)\right)}{\sin\pi J}\,.\label{eq:t_partial_wave_3}
\end{align}
The contour is next distorted as shown in Fig. \ref{fig:Complex_J}(b).  Assuming that there is at least one pole in the complex-J plane at $J=\alpha(t)$ introduces the closed contour $C''$ and assuming that the semicircular part of $C'$ vanishes at infinity gives
\begin{align}
M^{(a)IPG}_{\pm}(s,t)=&-\frac{1}{2i}\int_{-\frac{1}{2}-i\infty}^{-\frac{1}{2}+i\infty} dJ \frac{(2J+1)f^{IPG}_{\pm}(J,t)\left( P_J(-z)\pm P_J(z)\right)}{\sin\pi J}\nonumber\\
&+\frac{\pi(2\alpha(t)+1)\bar{\beta}^{IPG}_\pm(t)\left(P_{\alpha(t)}(-z)\pm P_{\alpha(t)}(z)\right)}{\sin\pi\alpha(t)}\,.\label{eq:t_partial_wave_4}
\end{align}
This is the Sommerfeld-Watson transform. The first term is the background integral which is unknown. The second term describes the Regge pole contribution where $\bar{\beta}^{IPG}_\pm(t)$ is the residue of $f^{IPG}_{\pm}(J,t)$  and $\alpha(t)$ is the trajectory of the Regge pole.

This expression can now be analytically continued to the s-channel cm frame which describes physical $NN$ scattering. In this frame $s\geq 4m^2$, $4m^2-s\leq t\leq 0$ and $z\geq 1$. For $z>0$,
\begin{equation}
P_\alpha(-z)=-\frac{2\sin\pi\alpha}{\pi}Q_\alpha(z)+e^{-i\pi\alpha}P_\alpha(z)
\end{equation}
And for $z>>1$ the asymptotic forms of the Legendre functions are given by
\begin{align}
P_\alpha(z)&\rightarrow\frac{2^\alpha\Gamma\left(\alpha+\frac{1}{2}\right)}{\sqrt{\pi}\Gamma(\alpha+1)}z^\alpha\\
Q_\alpha(z)&\rightarrow\frac{\sqrt{\pi}\Gamma(\alpha+1)}{2^{\alpha+1}\Gamma\left(\alpha+\frac{3}{2}\right)}z^{-(\alpha+1)}.
\end{align}
For large values of $z$, the background integral goes as $z^{-\frac{1}{2}}$. So this contribution can be suppressed by requiring that $z$ be large which implies that $s$ must be large compared to $4m^2$ and $|t|$ must be small. The Legendre function in the Regge pole contribution can also be expanded in $z$. Since the residue $\bar{\beta}^{IPG}_\pm(t)$ is undetermined, we can absorb all of the real $t$ dependent factors in the second term of (\ref{eq:t_partial_wave_4}) and those arising from the expansion of the Legendre functions into the residue function $\beta^{IPG}_\pm(t)$. The  Regge pole contributions necessary to determine the invariants $F^{I}(s,t)$ can be parameterized as
\begin{equation}\label{eq:ReggeExchange}
 R^{IPG}_{\pm j}(s,t)= \zeta(s,t)\sum_{k}\xi_{k\pm}(t)\beta_{\pm k}^{IPG}(t) z^{\alpha_k(t)},
\end{equation}
where $\beta^{IPG}(t)$ is the residue and $\xi_{\pm}(t)$ is a phase function. In fitting the Regge model to $NN$ scattering observables, we use three forms of the residue function given by
\begin{align} \label{EQ:residues}
\beta_{I}(t)   &= \beta_0e^{\beta_1t} \nonumber \\
\beta_{II}(t)  &= \left(1 - e^{\gamma t}\right)\beta_0e^{\beta_1t}  \\
\beta_{III}(t) &= \frac{t}{4m^2}\beta_0e^{\beta_1t} \nonumber
\end{align}
where $\beta_0$, $\beta_1$ and $\gamma$ are fitting parameters. The phase function is
\begin{equation}\label{eq:signature}
\xi_{\pm}(t) = \left\{ \begin{array}{cc}
 e^{-i(\pi\alpha(t)/2 + \delta)} &  + \\  e^{-i(\pi(\alpha(t)+1)/2 + \delta)} &  -
\end{array}  \right.
\end{equation}
where $\alpha(t)=\alpha_0+\alpha_1 t$, and $\delta$ is a phase that is a fitting parameter. The function $\zeta(s,t)$ is a cutoff factor defined as
\begin{equation}
	\zeta(s,t) = \left(1 - e^{20\left(\frac{t}{4m^2 - s} - 1\right)}\right)\,.
\end{equation}
This was introduced to improve the stability of the fitting procedure near $\theta=0^\circ$ and $180^\circ$. It has no effect at large $s$.

The five classes of Regge poles needed to produce the invariants $F^{I}_i(s,t)$ are
\begin{align}
R^{I}_1(s,t)&=R^{I++}_{+1}(s,t)\nonumber  \\
R^{I}_2(s,t)&=R^{I--}_{-2}(s,t)\nonumber  \\
R^{I}_3(s,t)&=R^{I+-}_{-3}(s,t)\nonumber  \\
R^{I}_4(s,t)&=R^{I--}_{+4}(s,t)\nonumber  \\
R^{I}_5(s,t)&=R^{I-+}_{+5}(s,t)
\end{align}
and the invariant functions are given by
\begin{equation}
F^{I}_i(s,t)=\Xi_{ij}(s,t)R^{I}_j(s,t)\,,
\end{equation}
where $i\in\{S,V,T,P,A\}$ and
\begin{align}
\Xi_{S1}(s,t)&=-\frac{m^2}{2(4m^2-t)}\nonumber  \\
\Xi_{V2}(s,t)&=-\frac{4m^2-t}{8\left(2s+t-4m^2 \right)}\nonumber  \\
\Xi_{V3}(s,t)&=\frac{t}{8(2s+t-4m^2)}\nonumber  \\
\Xi_{T3}(s,t)&=-\frac{m^2}{4(2s+t-4m^2)}\nonumber  \\
\Xi_{P4}(s,t)&=-\frac{m^2}{2t}\nonumber  \\
\Xi_{A5}(s,t)&=\frac{1}{8}\,.
\end{align}
All other components of the matrix are 0.

We also utilize a factor of $\frac{t}{4m^2}$ for type 4 exchanges, which was necessary to guarantee conservation of angular momentum.
Type 5 exchanges were multiplied by $\frac{4m^2}{s}$, which we assume we can factor from $F_{A}(s,t)$.
This is necessary in order to cancel with an additional factor of $s$, which occurs when performing the Dirac algebra to calculate the NN helicity amplitudes, and ensures unitarity for large $s$.

Parametrizing the Fermi invariants in terms of Regge poles provides a straightforward method of ensuring that Regge exchanges of appropriate quantum numbers contributes appropriately to the NN scattering. It also greatly simplifies the Regge analysis in that it reduces to the spinless case. While the techniques vary, conceptually similar approaches have been used in the past \cite{gribov,sharp}.

The above describes only the contributions of the direct diagram Fig. \ref{fig:symmetry}(a). The exchange contributions, Fig. \ref{fig:symmetry}(b), are obtained by the replacement $t\rightarrow u$ and a mixing of the invariants. The complete $pn$ invariants can then be written as
\begin{equation}
{\cal F}^{pn}_k(s,t)=F^{0}_k(s,t)-F^{1}_k(s,t)-2S_{kl}F^{1}_l(s,u)\,,
\end{equation}
where
\begin{equation}
\bm{S}=\left( \begin{array}{rrrrr}
\frac{1}{4} & 1 & 3 & \frac{1}{4} & -1 \\
\frac{1}{4} & -\frac{1}{2} & 0 & -\frac{1}{4} & -\frac{1}{2}\\
\frac{1}{8} & 0 & -\frac{1}{2} & \frac{1}{8} & 0 \\
\frac{1}{4} & -1 & 3 & \frac{1}{4} & 1 \\
-\frac{1}{4} & -\frac{1}{2} & 0 & \frac{1}{4} & -\frac{1}{2}
\end{array}\right)
\end{equation}
The complete $pp$ invariants are given by
\begin{equation}
{\cal F}^{pp}_k(s,t)= F^{0}_k(s,t)+F^{1}_k(s,t)-S_{kl}\left( F^{0}_l(s,u)+F^{1}_l(s,u)\right)\,.
\end{equation}
Our object was to construct a fit to $NN$ scattering observables that could be used at higher invariant masses than allowed by the SAID partial wave analysis. We were motivated to start with Regge theory as a result of its success in describing this process for large values of $s$ and then to connect it to the region of $5.4\ {\rm GeV^2}\leq s\leq 12\ {\rm GeV^2}$ where a considerable amount of data is available for a variety of observables. In doing this, we are clearly venturing into the region where many of the assumptions of Regge theory are no longer valid. Our approach was to retain the form of the Regge theory to provide fits in this region by introducing a set of ``effective'' trajectories with no relation to the meson trajectories to provide a basis for fitting in this region. The fit which we obtained was respectable given the wide range of $s$ and the large number of data sets produced at a variety of institutions over a long time period, but may obviously be subject to improvement. We will now compare the final state interactions contributions to deuteron electrodisintegration observables for this fit and that derived from the SAID helicity amplitudes.

\section{Results}

\begin{table}
\caption{The kinematics for the $x=1$ and $x=1.3$ kinematics used in the electrodisintegration calculations presented here.}
\begin{tabular}{crr}
\hline\hline
kinmatics & 1 &2\\\hline
$x$ & 1.0 & 1.3\\
$Q^2\ {\rm (GeV)^2}$& 2.4 & 4.5 \\
$E_{beam}\ {\rm (GeV)}$& 6.25& 8.6\\
$s\ {\rm (GeV)^2}$ & 5.91 & 5.93\\
$\theta_e\ {\rm (deg)}$&15.97& 16.0 \\
$\frac{Q^2}{q^2}$ & 0.60 & 0.57 \\
\hline\hline
\end{tabular}\label{tab:kinematics}
\end{table}

In this section we present a comparison to various observables for deuteron electrodisintegration for unpolarized hadrons, polarized deuteron and polarized final state proton. Two different kinematics are used, one for Bjorken x of $x=1.0$ and the other for $x=1.3$.  Various kinematical variables for the two kinematics are shown in Table \ref{tab:kinematics}. The kinematics are chosen such that $s$, the electron scattering angle $\theta_e$ and the ratio of the square of four-momentum transfer to three-momentum transfer $Q^2/q^2$ are approximately equal for the two cases. The values of $s$ are close to the upper range available from SAID and are at the lower end of the fitting range for the Regge parameterization. In all cases the onshell approximation for the final state interactions (FSI), as described in \cite{JVO_2008_newcalc}, is used. An offshell prescription for the SAID FSI was proposed in \cite{JVO_2008_newcalc}, but a more complete approach is possible for the Regge parameterization and will be considered in a future paper. Since both the SAID and Regge amplitudes are fit to onshell data, using only the onshell approximation is the most reasonable way to compare the two methods.

\begin{figure}
    \includegraphics[width=6in]{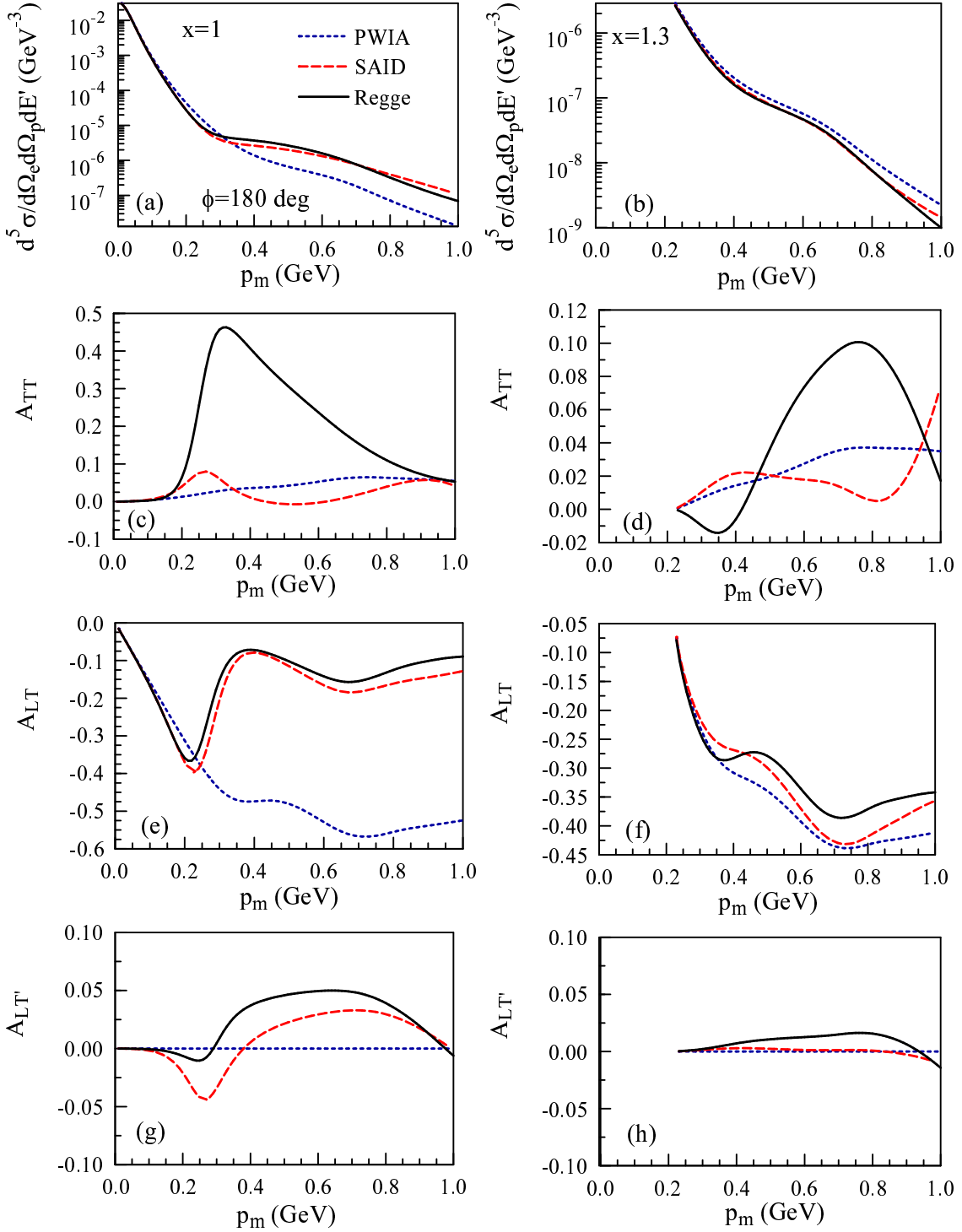}
    \caption{(color online) Spin observables for unpolarized hadrons. Short-dashed lines represent the PWIA contribution. Long-dashed lines include the SAID FSI and solid lines include the Regge FSI. Plots in the left-hand column are for the $x=1$ kinematics and plots in the right-hand column are for the $x=1.3$ kinematics.}
    \label{fig:stat_un}
\end{figure}

\begin{figure}
    \includegraphics[width=6in]{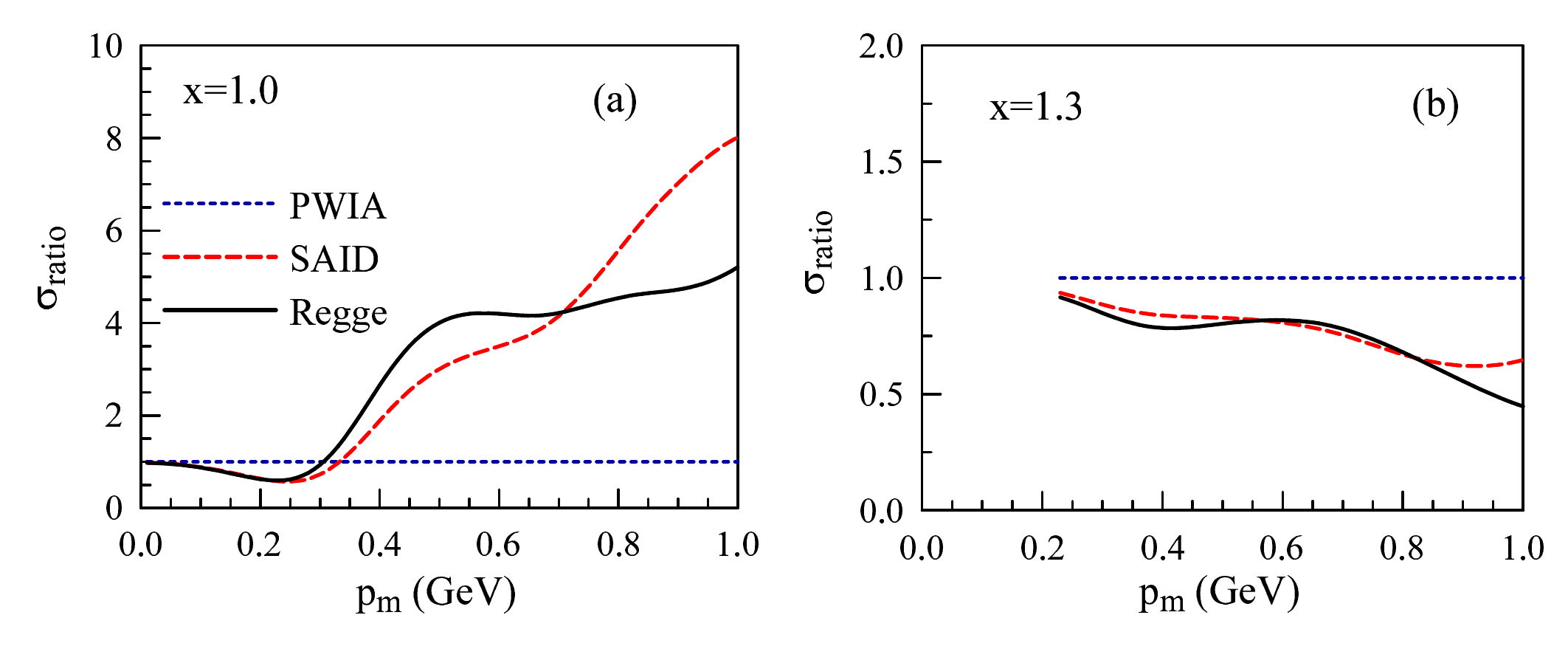}
    \caption{(color online) Ratios of the differential cross sections to the PWIA approximation. Lines are represented as in Fig. \ref{fig:stat_un}.}
    \label{fig:ratio}
\end{figure}

Figure \ref{fig:stat_un}  shows the observables for the case where neither the deuteron target nor the ejected proton are polarized, and where the azimuthal angle is chosen to be $\phi=180^\circ$. A summary of the various calculated quantities is contained in the appendix. Figures \ref{fig:stat_un}(a)and (b) show the differential cross sections as a function of missing momentum $p_m$ for the plane wave impulse approximation (PWIA) and for the SAID and Regge FSI for the $x=1$ kinematics and the $x=1.3$ kinematics respectively.
The size and shape of the two FSI calculations are similar in each case. Since these are semi-log plots, a more accurate evaluation of the differences is given by the ratio of distorted wave to PWIA cross section $\sigma_{ratio}$ as is shown in Fig. \ref{fig:ratio}.
In Fig. \ref{fig:ratio}(a) for the $x=1$ kinematics, the SAID and Regge results are very similar for $p_m<0.3\ {\rm GeV}$ but differ by up to 50 percent from the PWIA result. At higher missing momenta both the SAID and Regge results become increasingly large compared to the PWIA, and reach a value of approximately 8 times the PWIA at $p_m=1.0\ {\rm GeV}$ for the SAID FSI and approximately 5 times for the Regge FSI. For the $x=1.3$ kinematics, shown in Fig. \ref{fig:ratio}(b), the difference between the SAID and Regge FSI are much smaller and they are both much closer to the PWIA. Note that for $x=1$ both of the FSI lie above the PWIA but for $x=1.3$ they are below. This suggests that it may be possible to find kinematics at which the FSI effects are minimal and may allow for an approximate extraction of the deuteron ground-state momentum distribution, as has been suggested previously \cite{Boeglin_highmom}.

\begin{figure}
    \includegraphics[width=3in]{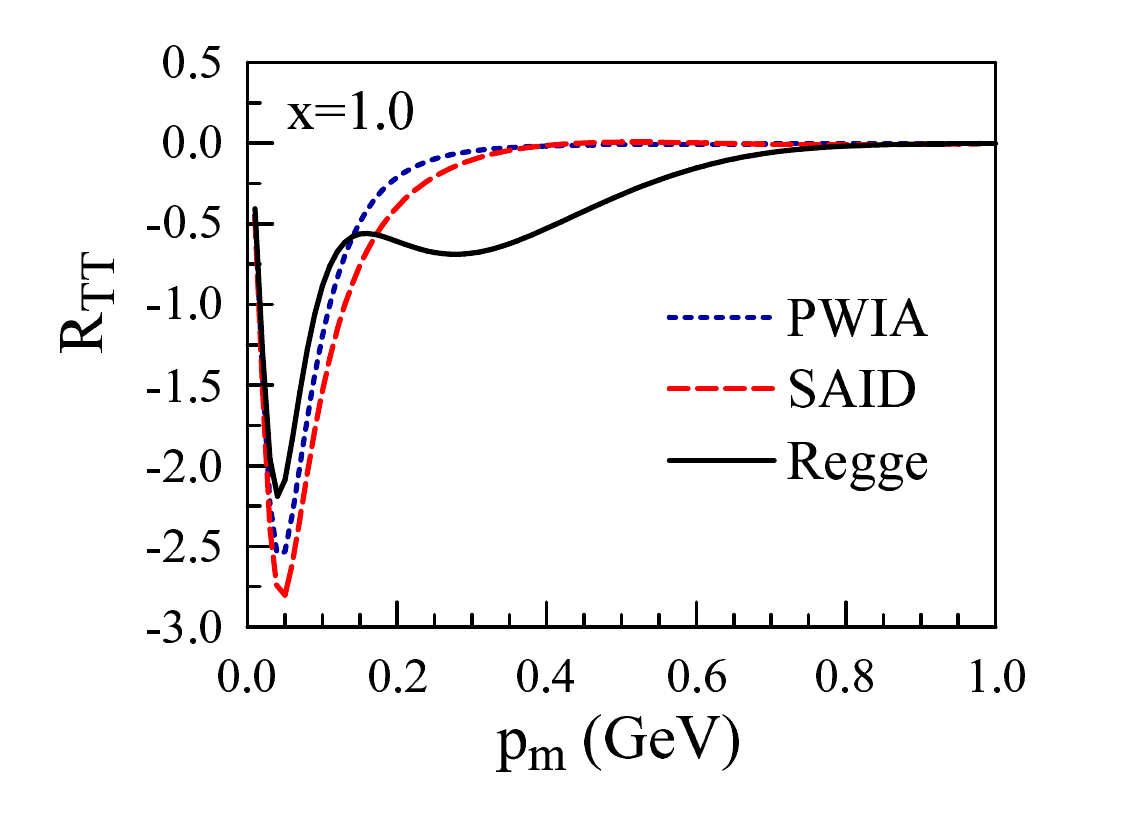}
    \caption{The transverse-transverse response function $R_{TT}$ for the $x=1$ kinematics. Lines are represented as in Fig. \ref{fig:stat_un}.}
    \label{fig:RTT}
\end{figure}

Figures \ref{fig:stat_un}(c) and (d) show the transverse-transverse asymmetry $A_{TT}$ for the $x=1$ and $x=1.3$ kinematics, respectively. This asymmetry, which is proportional to the $R_{TT}$ response function, is generally assumed to be small since $R_{TT}$ has generally been shown to be small. This is the case in \ref{fig:stat_un}(c) for the PWIA and SAID results, but the asymmetry for the Regge FSI is large for intermediate values of $p_m$. The reason for this can be seen from Fig. \ref{fig:RTT} which shows $R_{TT}$ for the $x=1$ kinematics.
Note that all three calculations minimum at around $p_m=0.05\ {\rm GeV}$ where the cross section is large. However, while the PWIA and Regge results fall smoothly to 0 with increasing $p_m$, the Regge results show a second minimum in a region where it is comparable in magnitude to the rapidly falling cross section. This results in the large values for $A_{TT}$ which involves a ratio of the transverse-transverse contribution to the cross section to the sum of the longitudinal and transverse contributions. It should be noted that the relationship between the Fermi invariants and the response functions is very complicated and can involve interferences between the various contributions. As a result we have not been able to isolate a single source for the second peak in the Regge $R_{TT}$ response function. Interference response functions and their associated asymmetries may show unpredictable sensitivities to small differences in the Fermi invariants. Ascertaining the significance of these differences requires that the errors in fitting parameters for the scattering amplitudes be propagated to the electrodisintegration calculations. This can be done for the Regge case since we can generate the hessian matrix for the fit. This will be done when sufficient resources are available. Unfortunately, we do not have access to similar information about the SAID helicity amplitudes.

Figures \ref{fig:stat_un}(e) and (f) show the longitudinal transverse asymmetry $A_{LT}$ for the $x=1$ and $x=1.3$ kinematics respectively. At $x=1$, this asymmetry is relatively large and the two FSI models give comparable results and differ substantially form the PWIA. At $x=1.3$, the two FSI models have similar form but tend to be in less agreement than in the $x=1$ case. Both, however, are much closer to the PWIA result.

Figures \ref{fig:stat_un}(g) and (h) show the longitudinal transverse asymmetry $A_{LT'}$ for the $x=1$ and $x=1.3$ kinematics respectively. Measurement of the asymmetry requires a polarized electron beam. Since this response is odd under the combination of time reversal and parity, its value is 0 in PWIA. For both kinematics it is small for both FSI models. The significance of the differences between the SAID and Regge results is unclear.

\begin{figure}
    \includegraphics[width=6in]{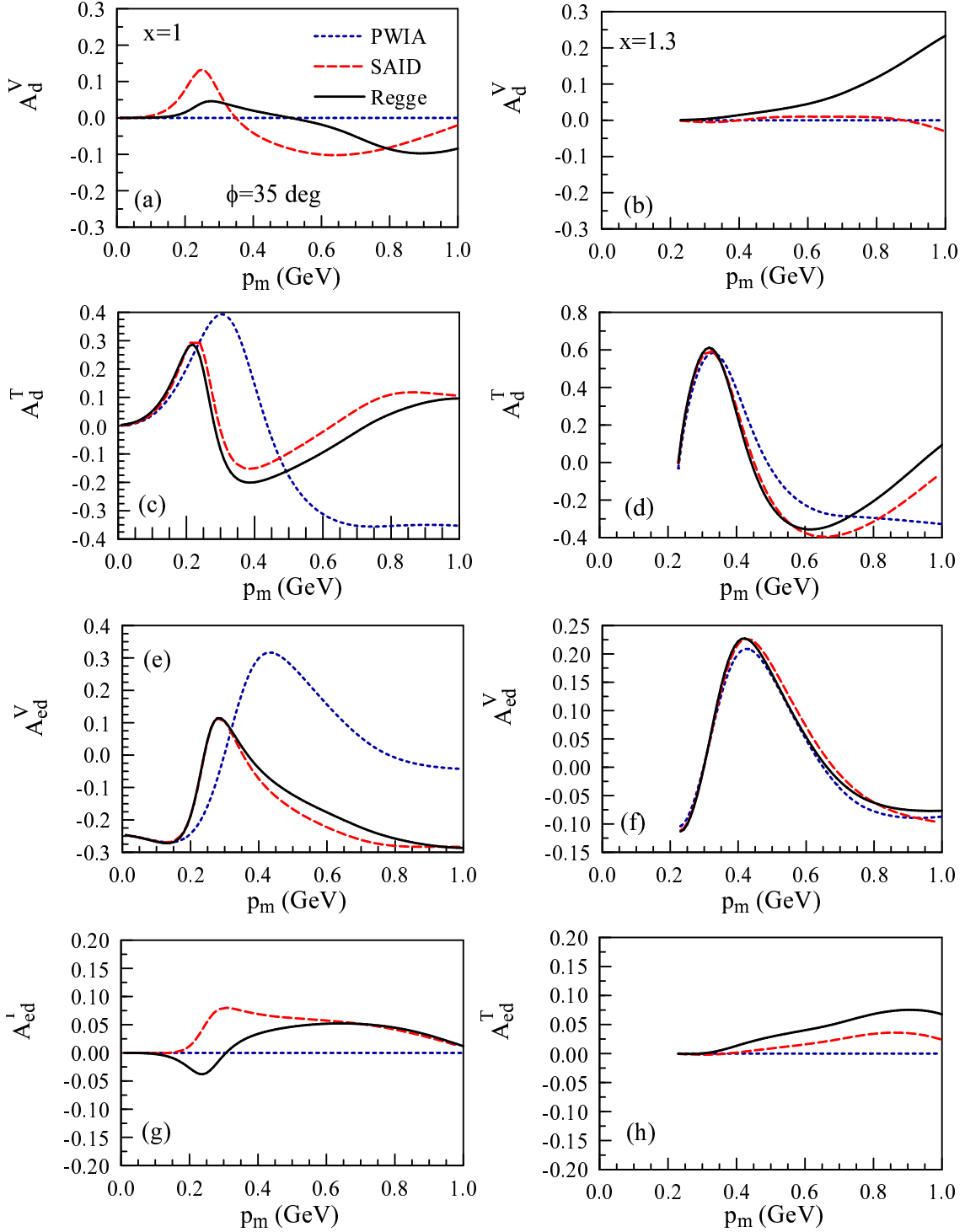}
     \caption{(color online) Single and double spin asymmetries for vector and tensor polarizations along the beam axis. Plots in the left-hand column are for the $x=1$ kinematics and plots in the right-hand column are for the $x=1.3$ kinematics. Lines are represent as in Fig. \ref{fig:stat_un}.}
    \label{fig:stat_targ}
\end{figure}

Figure \ref{fig:stat_targ} shows the single and double spin asymmetries for vector and tensor polarization of the target deuteron along the direction of the electron beam at an azimuthal angle of $\phi=35^\circ$. We see good agreement in these observables between the Regge and SAID approach as well as strong effects from the FSI. This suggests that target polarization asymmetries can provide insight to the effects of FSI while masking the model dependence of how these are calculated.

Figures \ref{fig:stat_targ}(a) and (b) shows the vector polarized target asymmetry $A^{V}_{d}$, for $x = 1$ and $x = 1.3$ kinematics respectively. Note that this asymmetry is zero in the absence of final state interactions. Qualitatively the Regge and SAID approaches are similar.

Figures \ref{fig:stat_targ}(c) and (d) show tensor polarized target asymmetry $A^{T}_{d}$, for $x = 1$ and $x = 1.3$ kinematics respectively. Here the two approaches are in excellent agreement and we see a dramatic change in behavior for $x = 1$. The FSI contributions to this observable are minimal at the $x = 1.3$ kinematics.

Figures \ref{fig:stat_targ} (e) and (f) show the double spin asymmetry for vector polarized target and polarized beam $A^{V}_{ed}$, for $x = 1$ and $x = 1.3$ kinematics respectively. We again note that the two approaches are in excellent agreement and observe that the FSI contributions are minimal at the $x = 1.3$.

Figures \ref{fig:stat_targ}(g) and (h) show the double spin asymmetry for tensor polarized target and polarized beam $A^{T}_{ed}$, for $x = 1$ and $x = 1.3$ kinematics respectively. This asymmetry is zero in the PWIA. Qualitatively the approaches yield similar results, and while the FSI do cause a non zero contribution the value is relatively small.

\begin{figure}
    \includegraphics[width=6in]{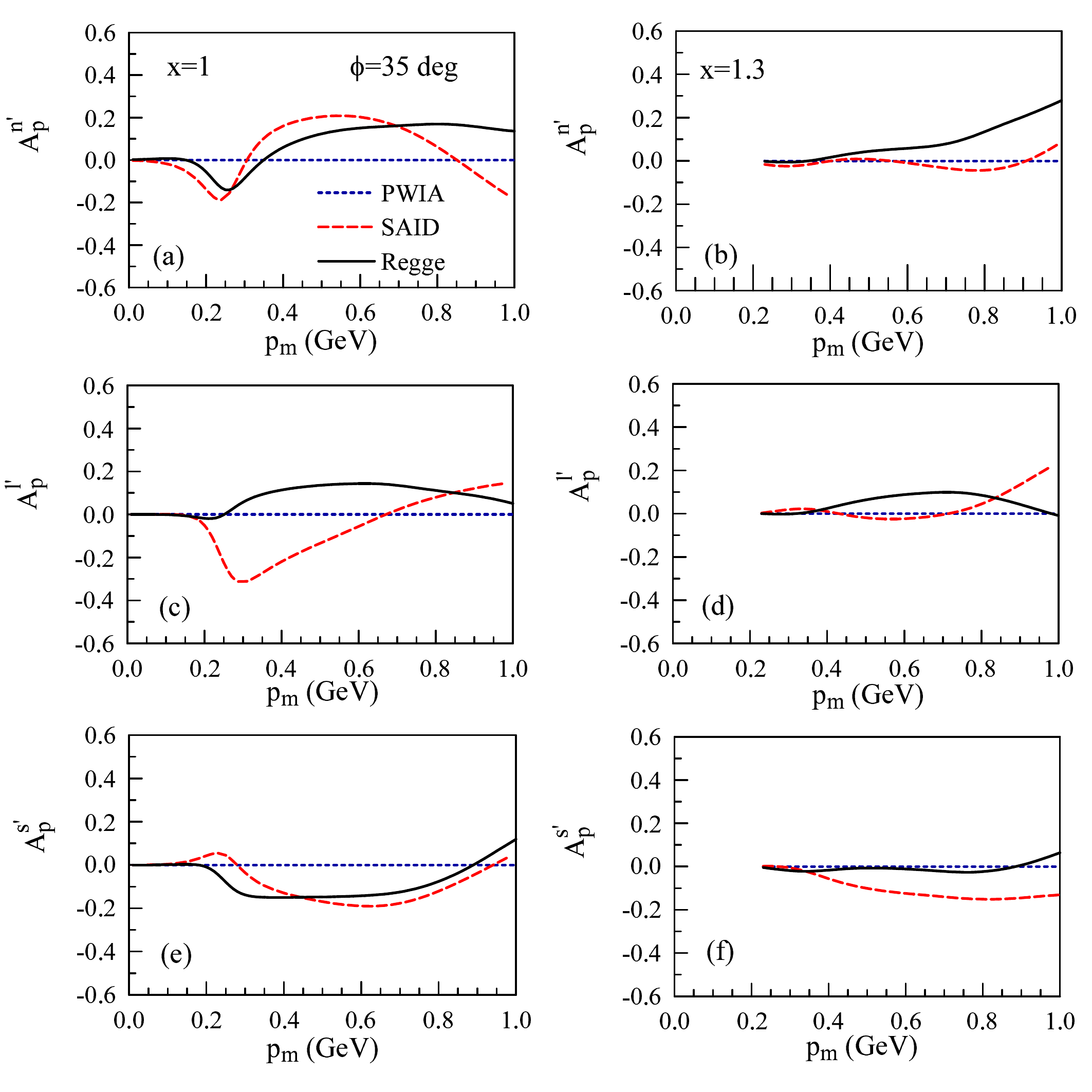}
    \caption{(color online) Single spin asymmetries for ejected protons polarized along the $\hat{n}'$, $\hat{l'}$ and $\hat{s}'$ directions. Plots in the left-hand column are for the $x=1$ kinematics and plots in the right-hand column are for the $x=1.3$ kinematics. Lines are represent as in Fig. \ref{fig:stat_un}.}
    \label{fig:stat_eject1}
\end{figure}

In Figure \ref{fig:stat_eject1} we present the results for polarized ejected proton at an azimuthal angle of $\phi=35^\circ$. All asymmetries in Figure \ref{fig:stat_eject1} are zero for PWIA, thus presenting an ideal set of asymmetries for exploring the contribution of FSI.

Figures \ref{fig:stat_eject1}(a) and (b) show the asymmetry $A^{n'}_{p}$, for $x = 1$ and $x = 1.3$ kinematics respectively, and we see good agreement between the models. Figures \ref{fig:stat_eject1}(c) and (d) show the asymmetry $A^{l'}_{p}$. For the $x = 1$ kinematics we observe that the model approaches are similar in magnitude but differ in sign. Because of the strong model dependence evident in this observable, and due to the relatively large value, this would provide an interesting measurement, which could shed light on the role of FSI as well as the various models used to calculate them. FSI effects at $x = 1.3$ are less pronounced. Figures \ref{fig:stat_eject1}(e) and (f) show the asymmetry $A^{s'}_{p}$. Here we again see good qualitative agreement between the two models.

\begin{figure}
    \includegraphics[width=6in]{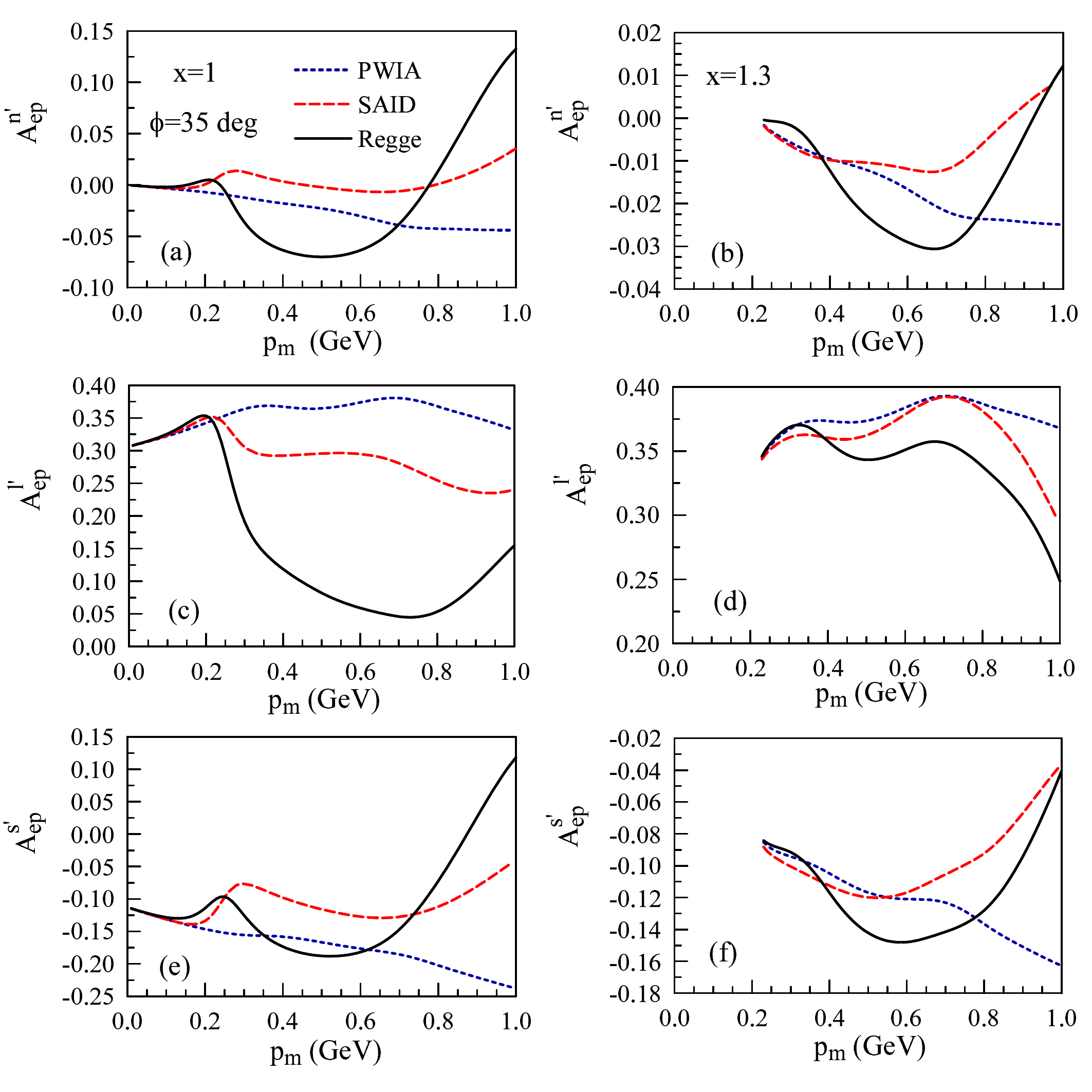}
    \caption{(color online) Double spin asymmetries for ejected protons polarized along the $\hat{n}'$, $\hat{l'}$ and $\hat{s}'$ directions. Plots in the left-hand column are for the $x=1$ kinematics and plots in the right-hand column are for the $x=1.3$ kinematics. Lines are represent as in Fig. \ref{fig:stat_un}.}
    \label{fig:stat_eject2}
\end{figure}

Figure \ref{fig:stat_eject2} shows the double spin asymmetries for polarized beam and polarized ejected proton at an azimuthal angle of $\phi=35^\circ$.
Figures \ref{fig:stat_eject2}(a) and (b) show the asymmetry $A^{n'}_{ep}$, for $x = 1$ and $x = 1.3$ kinematics respectively. Again we note that this asymmetry is highly sensitive to FSI model dependence at $x = 1$  causing deviation in opposite directions to the PWIA, although the magnitude of the deviation is relatively small. The same behavior is observed for $x = 1.3$, however less dramatic.

Figures \ref{fig:stat_eject2}(c) and (d) show the asymmetry $A^{l'}_{ep}$, for $x = 1$ and $x = 1.3$ kinematics respectively. In this case we observe qualitatively the same behavior between the SAID and Regge approaches although the Regge model is much more drastic at $x = 1$. At $x = 1.3$ we see similar, albeit less pronounced effects. Due to the large differences between the approaches we again point out that measurements of this asymmetry would prove insightful.

Figures \ref{fig:stat_eject2}(e) and (f) show the asymmetry $A^{s'}_{ep}$, for $x = 1$ and $x = 1.3$ kinematics respectively. We note that for both kinematics the two models are qualitatively similar, with relatively small deviations from the PWIA and each other.

\section{Conclusions}
In this paper we have presented a new method of calculating final state interactions for the electrodisintegration of the deuteron. The FSI are calculated using a Regge model, which was fit to available NN scattering data. The model is fully relativistic and incorporates full spin dependence. With the addition of this new method we have significantly extended the kinematic range of our deuteron electrodisintegration calculation which was previously limited due to the absence of high energy NN amplitudes from SAID. We have presented results for kinematic regions where the SAID and Regge approaches overlap for comparison.

The comparisons suggest that for most of the observables there is good agreement between the two approaches. We expect that most discrepancies between the two models would be within error bands were they available. We anticipate being able to propagate the error for the Regge model once sufficient resources are available, however, an error analysis requires information that is unavailable from SAID. Propagation of the $NN$ fitting error to the electrodisintegration observables will require a substantial amount of computational resources.

The results are consistent with expectations that FSI play a vital role in understanding the reaction mechanism. We have noted that from our results there may be kinematic regions where FSI are minimized and PWIA is a valid approximation. This is most evident in the cross section and polarized target asymmetries. We have also identified asymmetries with large FSI contribution and significant sensitivity to the model dependence of the two approaches. In particular $A_{TT}$, $A^{l'}_{p}$ and $A^{l'}_{ep}$. While almost all observables are sensitive to FSI and measurements would prove useful, these are particularly interesting because of the discrepancies between the two models.

Now that we have tested our Regge model in the region of overlap available with SAID we anticipate future work. Offshell effects should be taken into account and the Regge model allows for a natural offshell extrapolation. In addition we now have the capabilities to explore the kinematic region which will become accessible once the Jefferson Lab 12 GeV upgrade is completed.

{\bf Acknowledgments}:   This work was
supported in part by funds provided by the U.S. Department of Energy
(DOE) under cooperative research agreement under No.
DE-AC05-84ER40150 and by the National Science Foundation under
grant No. PHY-1002478.

\appendix
\section{Deuteron Electrodisintegration Observables}

\begin{figure}
\centerline{\includegraphics[height=3in]{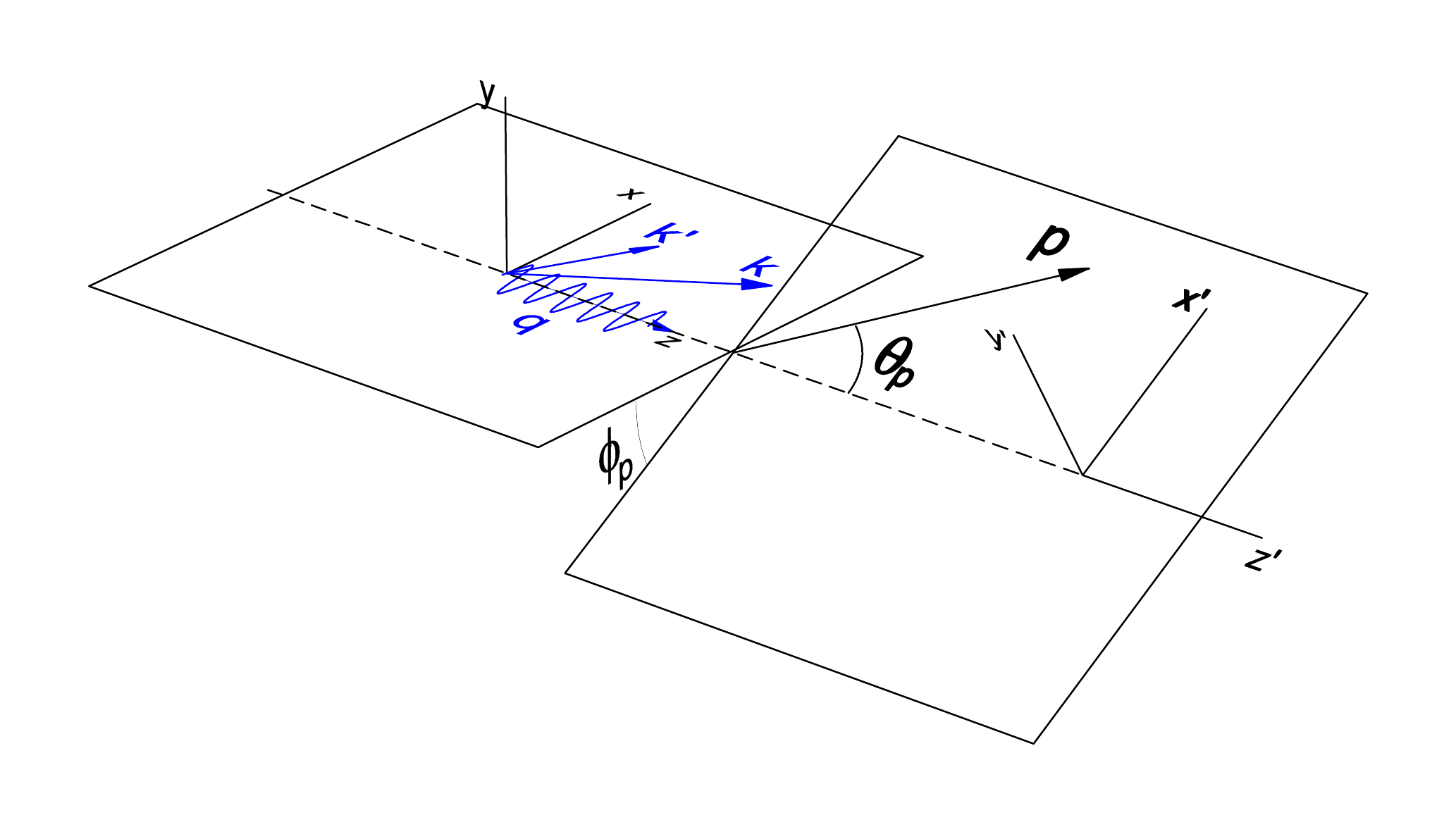}}
\caption{(Color online) Coordinate systems for the $D(e,e'p)$ reaction.  $k$ and $k'$ are the initial and final electron four-momenta,
$q$ is the four-momentum of the virtual photon and $p$ is the four-momentum of the final-state proton.  }\label{coordinates}
\end{figure}

The hadronic response tensor for deuteron electrodisintegration with polarization of the target deuteron and the ejected proton can be written as
\begin{align}
w_{\lambda'_\gamma,\lambda_\gamma}(D,\hat{\cal S})=&\sum_{s_1,s'_1,s_2,\lambda_d,\lambda'_d}
\left<\bm{p}_1s'_1;\bm{p}_2s_2;(-)\right|J_{\lambda'_\gamma}\left|\bm{P}\lambda'_d\right>^*
\left<\bm{p}_1s_1;\bm{p}_2s_2;(-)\right|J_{\lambda_\gamma}\left|\bm{P}\lambda_d\right>\nonumber\\
&\rho^D_{\lambda_d\lambda'_d}(D)\rho^{p}_{s_1s_1'}(\hat{\cal S})
\end{align}
where
\begin{equation}
J_{\pm 1}=\mp\frac{1}{\sqrt{2}}(J^1\pm J^2)
\end{equation}
and
\begin{equation}
J_0=J^0\,.
\end{equation}
The deuteron density matrix $\rho^D_{\lambda_d\lambda'_d}(D)$ is a function of
\begin{equation}
D\in\{U,T_{10},T_{11},T_{20},T_{21},T_{22}\}\,,
\end{equation}
where $U$ designates an unpolarized deuteron and the $T_{ij}$ are the components of the polarization tensor as described in more detail in \cite{JVO_2009_tar_pol}.
The proton density matrix is given by
\begin{equation}
\rho^{p}_{s_1s_1'}(\hat{\cal S})=\frac{1}{2}\chi^\dagger_{s_1}({\bf 1}+{\bm \sigma}\cdot\hat{\cal S})\chi_{s'_1}\,.
\end{equation}
where $\hat{\cal S}$ is a unit vector in the direction of the proton polariation it its rest frame.

The general form of the $D(e,e'p)$ cross section can be written in the lab frame as
\cite{raskintwd,dmtrgross}

\begin{eqnarray}
\left ( \frac{ d \sigma^5}{d E' d \Omega_e d \Omega_p} \right
)_{h,D,\hat{\cal S}}  & = & \frac{m_p \, m_n \, p_p}{16 \pi^3 \, M_d} \,
\sigma_{Mott} \,
f_{rec}^{-1} \,
 \Big[  v_L{\cal R}_L +   v_T {\cal R}_T
 + v_{TT} {\cal R}_{TT} + v_{LT}{\cal R}_{LT}
  \nonumber \\
& & +  h \,  v_{LT'} {\cal R}_{LT'}+h\,v_{T'}{\cal R}_{T'}
\Big] \, , \label{xsdef}
\end{eqnarray}
where $M_d$, $m_p$ and $m_n$  are the masses of the deuteron, proton and neutron,
 $p_p=p_1$ and $\Omega_p$
are the momentum and solid angle of the ejected proton, $E'$ is the
energy of the detected electron and $\Omega_e$ is its solid angle, with
$h=\pm 1$  for positive and negative electron helicity. The Mott cross
section is
\begin{equation}
\sigma_{Mott} = \left ( \frac{ \alpha \cos(\theta_e/2)} {2
\varepsilon \sin ^2(\theta_e/2)} \right )^2
\end{equation}
and the recoil factor is given by
\begin{equation}
f_{rec} = \left| 1+ \frac{\omega p_p - E_p q \cos \theta_p} {M_d \, p_p}
\right| \, . \label{defrecoil}
\end{equation}
The leptonic coefficients $v_K$ are
\begin{eqnarray}
v_L&=&\frac{Q^4}{q^4}\\
v_T&=&\frac{Q^2}{2q^2}+\tan^2\frac{\theta_e}{2}\\
v_{TT}&=&-\frac{Q^2}{2q^2}\\
v_{LT}&=&-\frac{Q^2}{\sqrt{2}q^2}\sqrt{\frac{Q^2}{q^2}+\tan^2\frac{\theta_e}{2}}\\
v_{LT'}&=&-\frac{Q^2}{\sqrt{2}q^2}\tan\frac{\theta_e}{2}\\
v_{T'}&=&\tan\frac{\theta_e}{2}\sqrt{\frac{Q^2}{q^2}+\tan^2\frac{\theta_e}{2}}
\end{eqnarray}

The response functions are related to the hadronic tensor by
\begin{align}
{\cal R}_L(D,\hat{\cal S})&=w_{00}(D,\hat{\cal S})\nonumber\\
{\cal R}_T(D,\hat{\cal S})&=w_{11}(D,\hat{\cal S})+w_{-1-1}(D,\hat{\cal S})\nonumber\\
{\cal R}_{TT}(D,\hat{\cal S})&=2\Re(w_{1-1}(D,\hat{\cal S}))\nonumber\\
{\cal R}_{LT}(D,\hat{\cal S})&=-2\Re(w_{01}(D,\hat{\cal S})-w_{0-1}(D,\hat{\cal S}))\nonumber\\
{\cal R}_{LT'}(D,\hat{\cal S})&=-2\Re(w_{01}(D,\hat{\cal S})+w_{0-1}(D,\hat{\cal S}))\nonumber\\
{\cal R}_{T'}(D,\hat{\cal S})&=w_{11}(D,\hat{\cal S})-w_{-1-1}(D,\hat{\cal S})\,.
\label{respdefxyz}
\end{align}

\subsection{Unpolarized}

In the case where the deuteron and proton are not polarized the response functions can be written as
\begin{align}
{\cal R}_L(U,\hat{0})&=R_L\nonumber\\
{\cal R}_T(U,\hat{0})&=R_T\nonumber\\
{\cal R}_{TT}(U,\hat{0})&=R_{TT}\cos 2\phi\nonumber\\
{\cal R}_{LT}(U,\hat{0})&=R_{LT}\cos\phi\nonumber\\
{\cal R}_{LT'}(U,\hat{0})&=R_{LT'}\sin\phi\nonumber\\
{\cal R}_{T'}(U,\hat{0})&=0
\end{align}
where the dependence on the azimuthal angle is written explicitly and the response functions $R_i$ are independent of this angle.

Three asymmetries can be defined for the interference response functions,

\begin{equation}
A_{TT}=\frac{v_{TT}R_{TT}}{v_L R_L+v_T R_T}\label{defatt}\,,
\end{equation}

\begin{equation}
A_{LT}= \frac{v_{LT}R_{LT}}{v_L R_L+v_T R_T+v_{TT}R_{TT}}\label{defalt}
\end{equation}
and
\begin{equation}
\displaystyle{ A_{LT'} =   \frac{v_{LT'} R_{LT'}}{v_L R_L + v_T R_T -
v_{TT} R_{TT} }} \label{defatlp}  \,,
\end{equation}
Note that while $A_{LT}$ can be obtained by measuring protons in the electron scattering plane symmetrically about the direction of the three-momentum transfer, the asymmetries $A_{TT}$ and $A_{LT'}$ require measurements to be made out of the scattering plane.
The asymmetry $A_{LT'}$ is defined as an electron single spin asymmetry and can, therefore, be easily obtained by flipping the beam helicity. While $R_L$ and $R_T$ are independent of photon-helicity-dependent phases, the interference response functions are not.  As a result, the interference response functions can be very sensitive to phase differences generated by non-nucleonic currents and final state interactions.  This is particularly true of $R_{LT'}$ which can be shown to be zero in the PWIA. The interference
response function, $R_{LT}$, is very sensitive to the relativity included in
the current operator, due to the various interference contributions
from the charge and transverse current operators \cite{relcur1,relcur2}.

\subsection{Target Polarization}

It is usually assumed that electron scattering cross sections are polarized along the direction of the three-momentum transfer $\bm{q}$. In this case it is possible to express the cross section in a form where the explicit $\phi$ dependence can be factored out of the response functions. In the case of a polarized target, it is often more convenient experimentally to polarize the target along the direction of the electron beam. The procedure for performing the appropriate rotations to obtain the cross section for this situation are described in \cite{JVO_2009_tar_pol}. This is used in the calculations presented in the paper. We present asymmetries for vector and tensor polarization along the beam axis.
The single and double asymmetries for these two polarizations are defined as
\begin{eqnarray}
\label{asymdef}
A^V_d&=&\frac{v_L {\cal R}_L(\widetilde{T}_{10},\hat{0})+v_T {\cal R}_T(\widetilde{T}_{10},\hat{0})+v_{TT}{\cal R}_{TT}(\widetilde{T}_{10},\hat{0})+v_{LT} {\cal R}_{LT}(\widetilde{T}_{10},\hat{0})}{\widetilde{T}_{10}\Sigma}\nonumber\\
A^T_d&=&\frac{v_L {\cal R}_L(\widetilde{T}_{20},\hat{0})+v_T {\cal R}_T(\widetilde{T}_{20},\hat{0})+v_{TT}{\cal R}_{TT}(\widetilde{T}_{20},\hat{0})+v_{LT} {\cal R}_{LT}(\widetilde{T}_{20},\hat{0})}{\widetilde{T}_{20}\Sigma}\nonumber\\
A^V_{ed}&=&\frac{v_{LT'}{\cal R}_{LT'}(\widetilde{T}_{10},\hat{0})
+v_{T'}{\cal R}_{T'}(\widetilde{T}_{10},\hat{0})}{\widetilde{T}_{10}\Sigma}\nonumber\\
A^T_{ed}&=&\frac{v_{LT'}{\cal R}_{LT'}(\widetilde{T}_{20},\hat{0})+v_{T'}{\cal R}_{T'}(\widetilde{T}_{20},\hat{0})
}{\widetilde{T}_{20}\Sigma}
\end{eqnarray}
where
\begin{equation}
\Sigma=v_L {\cal R}_L(U,\hat{0})+v_T {\cal R}_T(U,\hat{0})+v_{TT}{\cal R}_{TT}(U,\hat{0})+v_{LT} {\cal R}_{LT}(U,\hat{0})\,.
\end{equation}
 Here $R_i(\widetilde{T}_{10},\hat{0})$ and $R_i(\widetilde{T}_{20},\hat{0})$ denote the response functions where only $\widetilde{T}_{10}$ is nonzero or only $\widetilde{T}_{20}$ is nonzero. $R_i(U,\hat{0})$ denotes the unpolarized response functions.

 \subsection{Polarized Proton}

For unpolarized deuterons and polarized ejected protons, the proton polarization is typically described in terms of three unit vectors
\begin{align}
\hat{l}&=\frac{\bm{p}}{|\bm{p}|}\\
\hat{n}&=\frac{\bm{q}\times\hat{l}}{|\bm{q}\times\hat{l}|}\\
\hat{s}&=\hat{n}\times\hat{l}\,.
\end{align}
Using this coordinate system allows factorization of the response functions to display explicit dependence on $\phi$. However, defining asymmetries using this coordinate system produces asymmetries which are not properly defined for $\theta_p=0^\circ$ and $180^\circ$. An alternate set of basis vectors defined as
\begin{align}
\hat{l}'&=\hat{l}\\
\hat{s}'&=\frac{\hat{y}\times\hat{l}}{|\hat{y}\times\hat{l}|}\\
\hat{n}'&=\hat{l}'\times\hat{s}'
\end{align}
eliminates this problem but results in response functions that no longer factor to give the explicit $\phi$ dependence.

For convenience, define
\begin{equation}
\left ( \frac{ d \sigma^5}{d E' d \Omega_e d \Omega_p} \right
)_{h,U,\hat{\cal S}} = \sigma(\hat{\cal S})+h \sigma_h(\hat{\cal S})\,
\end{equation}
where $\sigma(\hat{\cal S})$ is the contribution to the cross section independent of the electron helicity and $\sigma_h(\hat{\cal S})$ is the part of the cross section proportional to $h$, the single and double asymmetries are now defined as
\begin{equation}
\label{asymdef1}
A^\xi_p=\frac{\sigma(\hat{\xi})}{\sigma(0)}
\end{equation}
and
\begin{equation}
\label{asymdef2}
A^\xi_{ep}=\frac{\sigma_h(\xi)}{\sigma(0)}
\end{equation}
where $\xi=n',l',s'$.

\bibliography{Regge_Deep.bib}
\end{document}